# Twenty years of experimental and numerical studies on microwave-assisted breakage of rocks and minerals—a review

Khashayar Teimoori[1] and Ferri Hassani[2]

[1,2]Department of Mining and Materials Engineering, McGill University, 3450 University, Frank Dawson Adams Bldg., Montreal, QC H3A 0E8, Canada

Abstract

Microwaves have been used for a variety of applications in the past two decades. However, there has been a significant and growing interest in the applications of microwaves in hard rock breakage and mineral processing industries. The purpose of this review paper is to focus on these applications and to present a careful review of the state-of-the-art experimental and numerical modeling techniques introduced in the literature from 2000 to 2020. The challenges involved in this research area are surveyed, and the efforts that should be made regarding the potential practical implementation of microwaves in industry are discussed.

## 1. Introduction

Over the years, mining companies have increased production rates due to increasing demand for raw materials from rapid growth in the global population and industrialization [1]. The mining industry is thus essential in order to ensure enough supply to support expansion. This expansion, improving production rates and reducing the footprint of mining operations, brings new opportunities to the mining industry's development since necessity leads to major investment in new innovative technologies. While today rock breakage is considered the most common operation in the civil and mining industries, there exist several challenges regarding the breakage of rocks in both rocks made up of different minerals and the mechanical cutting tools applied to them [2]. One of the main issues that a mining project might encounter during rock breakage processes is that the typical excavation machines available are operating at a limited level due to high disc wear and low rate of penetration in hard rocks [3,4]. Conventional rock breakage methods such as drilling, blasting, ripping, jack hammers, explosives, hydro-demolition, and heating through conduction produce high levels of noise, dust and/or vibration, which considerably affect the environment [5].

[E]-*mail addresses:* [1] Khashayar.teimoori@mail.mcgill.ca, Teimoori@usa.com (K. Teimoori).

[2] Ferri.hassani@mcgill.ca (F. Hassani)

Consequently, there has been significant industrial interest in replacing conventional methods of rock breakage with novel systems, as the bigger picture also considered the potential of creating non-explosive continuous rock breakage operations [6–8]. These new systems are expected to be environmentally friendly and to enhance levels of safety and productivity. More importantly, the proposed alternatives should be feasible and cost-effective [9]. They must achieve better rock stability, reduce induced seismicity due to the existence of high stresses, and maximize the extraction rate of the mechanical cutting tools. In this case, enormous efforts have been made in finding alternatives to reduce rocks' strength by affecting their mechanical and physical material properties prior to the use of mechanical tools [10–13]. This process of reducing rocks' strength is called "pre-conditioning". Various applications such as hydraulic splitting of rock, plasma blasting, thermal heating, chemical expansion powders, lasers, and electromagnetic induction (microwaves) have been introduced for rock pre-conditioning in recent decades [14]. The technology of microwave treatment of rocks has recently been assessed as a potential approach for rock pre-conditioning prior to breakage by mechanical means such as the Tunnel Boring Machine (TBM) [15,16]. It has been shown that the total energy applied to the rocks could be decreased by combining microwave energy and one of the available excavation machines, e.g. Roadheader, TBM, or Oscillating Disc Cutter (ODC). The penetration rate also increases when an excavation machine is equipped with microwave technology [17,18].

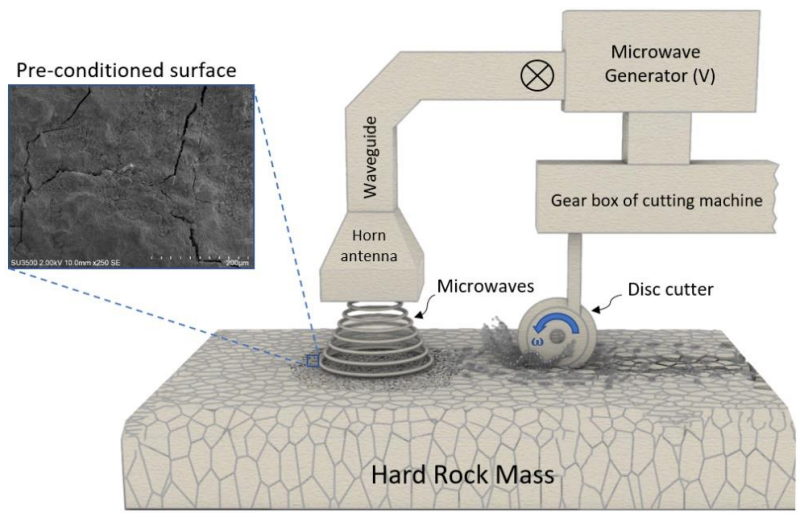

**Fig. 1.** A conceptual illustration of microwave-assisted rock pre-conditioning system prior to breakage by a disc cutter

The idea of the employing microwaves was first widely known by Maurer [19,20] as a novel application to break natural rocks with no mechanical tools involved. Later in 1991, Lindroth et al. [21] introduced the concept of rock pre-conditioning by microwaves prior to the use of mechanical discs as a potential alternative to conventional methods of rock breakage. Thereafter, several researchers have introduced the application of microwaves in the pre-conditioning of rocks as a novel technique in the future rock breakage operations [11,22–25]. A conceptual illustration of the primary idea of microwave-assisted rock pre-conditioning system prior to breakage by mechanical tools (e.g. TBM, disc cutter, Roadheader) is depicted in Fig.1.

Since the 2000s, microwave-assisted rock heating and pre-conditioning have been found to be a potential new avenue to improve mining and civil engineering projects related to breakage of rocks and processing of minerals [23,26]. The purpose of microwave treatment of rocks is to reduce their strength by weakening key material properties in a continuous manner and without explosives prior to the use of mechanical tools in order to facilitate the breakage operation [27–30]. In general, microwaves involve two different types of applicators (cavities), single-mode and multi-mode, with different heating and pre-conditioning mechanisms. Single-mode microwave irradiation is typically used for localized microwave effects (heating and cracking) on the surface of materials under irradiation; it is more applicable when (a) coupled with one of the available rock cutting tools such as drill and borehole, and/or (b) rock fragmentations needed for downstream processes [31–34]. On the other hand, multi-mode microwave irradiation is used for volumetric heating and pre-conditioning of rocks; it is more applicable in the mineral processing industry [35]. Thus, an understanding of different mechanisms involved in the processes of rock heating and pre-conditioning by the two types of microwaves is important for successful and efficient implementation of a microwave-assisted rock breakage or mineral processing system. To this end, many other parameters should be accurately considered and evaluated to arrive at a proper design prior to its being economically viable, since the only use of microwave technology to break rocks without mechanical tools requires a large amount of energy [36]. It has been shown that the effect of microwaves on rocks composed of different minerals is highly affected by variations in thermo-physical and electrical properties [30,37]. However, among different properties, the rock's dielectric properties, which comprise both dielectric constant and loss factor, play an important role in the amount of heating and damage produced after irradiation [38,39]. Today, there exist several challenges regarding the efficiency of microwaves that have to be resolved and properly addressed. The results contribute to better evaluating the impact of microwave technology on future rock breakage techniques, especially the energy-saving potential of such technologies, in both the mining and the processing of rocks. In terms of modeling and simulation, the present research project provides tools and insights into the design and performance of microwave-assisted rock breakage systems.

## 2. Evolution of microwave applications in rock mechanics and rock engineering

Even though electromagnetic waves were discovered in the mid-1800s, microwaves were first predicted by James Clerk Maxwell in 1864. Later in 1888, Heinrich Hertz proved the existence of microwaves by building a device that produced and detected microwave radiation [40]. Microwaves comprising electric and magnetic fields, represent a form of electromagnetic energy. Materials that can absorb microwave radiation are called dielectrics and contain dipoles. When a dielectric material is subjected to microwaves, its dipoles align and flip when the applied electromagnetic waves alternate. As a result, the stored internal energy is lost to friction and the dielectric materials heat up [41]. Microwaves have been used for a variety of applications in the past six decades. These applications include food processing, power transmission, communications, weather control, medical science, vulcanization of rubber, induction-based communication microwaves for Internet of Thing (IoT), medical sensors for exchanging information, as well as heating and drying [42–48]. However, among different applications of microwaves, heating of materials has been considered one of the most common applications of microwaves for domestic and industrial purposes [49,50], which therefore led to increased research into the applications of microwave irradiation in rock mechanics and the rock engineering field. The use of microwaves was first proposed by Maurer [19,20] as a novel application to break natural rocks with no mechanical tools involved. Maurer [19] illustrated the potential of some novel conceptual drilling techniques, of which one was a microwave-assisted drill. In short, the method that Maurer [19] introduced was drilling by only using microwave antenna at the end of the drill bit (see Fig. 2(a)). Although Maurer's [19] design was novel, many parameters had to be accurately considered in a proper design of a microwave-assisted rock drilling system to be economically viable, since the sole use of microwave technology to break rocks without mechanical tools requires a large amount of energy. The pressure to remain with the status quo always conserves the main focus in any technological change [51].

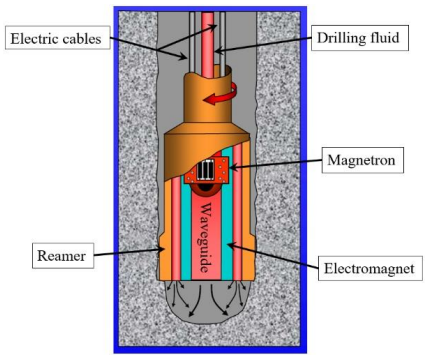  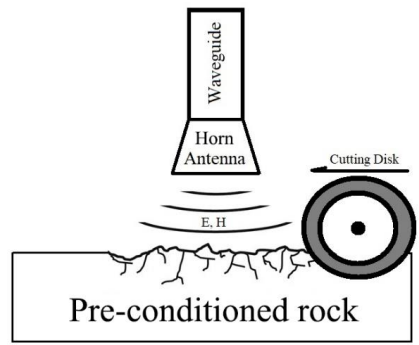

(a) modified from [19]                    (b) modified from Lindroth et al. [52]

**Fig. 2.** The primary idea of (a) the microwave-assisted drilling system, and (b) the microwave-assisted mechanical rock cutting system

After rock breakage, the ore needs to be processed in order to provide a valuable concentrate [53,54]. If the reduction in the wear on the drills is offset by determinantal effects on the processing, then new protocols for the separation of valuable minerals from the waste will be required [55]. Previous work investigating the effects of microwave irradiation on the processing of minerals, including separation, grinding, and purification, has shown some promise results [26,56–60]. From 1988 to 1991, it was discovered that ore minerals are great absorbers and gangue minerals are transparent to electromagnetic waves [61]. It has been shown that under microwave irradiation, the differential heating or selective heating is created between minerals grain and surrounding transparent matrix (gangue minerals), which generate intergranular cracks through minerals' boundaries. Moreover, the thermal expansion of different minerals causes tensile stress along the grain boundaries, which finally results in several microcracks. These microcracks reduce the overall rock strength and improve rock breakage, grindability, and mineral liberation. More importantly, the pre-conditioned rock facilitates rock drilling, reduces bit wearing, and decreases energy consumption of size reduction in mineral processing (e.g. crushing and grinding) [61]. These findings have been key focuses, since energy consumption and mineral recovery have always been two important issues in the mineral processing industry [62]. The very first recorded attempt to expose minerals to microwave irradiation was a patented work by Zavitsanos and Bleiler [63] in 1978 called "desulphurization of coal using microwaves". Subsequent research about this novel application of microwave irradiation on minerals was stimulated by a publication from Chen et al. [64] concerning the relative transparency of minerals to microwave energy. The research on application of microwaves for the pre-conditioning and breakage of rocks was first introduced in 1991 by Lindroth et al. [52] and called "pre-conditioning of rocks by microwaves" as a potential alternative to conventional methods. Following this innovation, the technology of microwave-assisted rock breakage was considered potentially as one of the most important explosive-free technologies for pre-conditioning of rocks prior to breakage by a cutting machine. A schematic of the first proposed novel microwave-assisted mechanical rock cutting system is shown in Fig. 2(b).

In 1994, the American National Research Council Committee on Advanced Drilling Technologies (CADT) published a report entitled "Drilling and Excavation Technologies for the Future" [65]. In this report, Maurer's [19] proposal on the use of the microwave-assisted drill was emphasized by the committee members. This report therefore underlined the importance of improved conventional mechanical drilling technologies as well as avenues to assist such technologies with the use of microwave energy. Thereafter, several researchers introduced the use of microwaves in the pre-conditioning and breakage of rocks and processing of minerals as a novel approach in the future mining operations. Extensive research began in the 2000s. Several studies have outlined that the total energy applied to the rocks can be decreased by combining microwave energy and one of the available excavation machines, e.g. Roadheader, TBM, or Oscillating Disc Cutter (ODC) [66,67]. The penetration rate also increases when an excavation machine is equipped with microwave technology [17]. In 2000, Jerby and Dikhtiar [68] presented their novel apparatus that drives the microwave energy from a 1 kW magnetron to a waveguide and, finally, into the material. The design is such that the applied microwaves are localized into a predetermined spot on the material. An extended project was later presented in 2003 which introduced a novel design of a Microwave Drill (MWD) system based on generating a hot spot on the surface of an object [69].

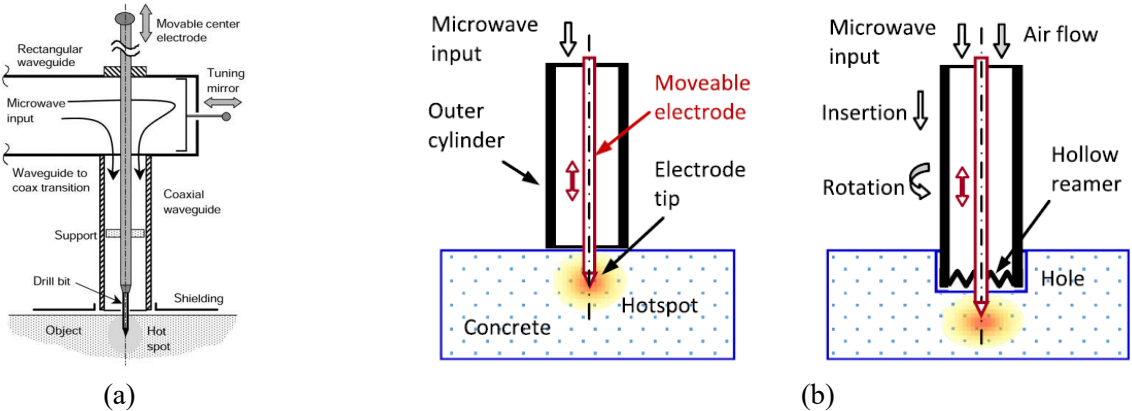

**Fig. 3.** (a) Scheme illustrating the principle underlying the operation of microwave-drill system [69]. (b) Conceptual schemes of MWD applicators, including a basic MWD consists of a coaxial waveguide with a moveable center electrode, inserted into the softened hot spot to form the hole (left) and an advanced MWD for deeper holes (right) [70]

As shown in Fig. 3(a), microwaves from a rectangular waveguide are transmitted to a coaxial waveguide. The central electrode (central antenna), which also function as a drill bit, concentrated the microwaves to a specific point. Therefore, the hot spot created on the surface weakened the rock's strength and resulted in easier drilling. Following the design presented in 2003, Jerby and colleagues developed the basic design of a microwave drill capable of drilling holes with 26 cm (depth) by 12 mm (diameter) in concrete, as shown in Fig. 3(b). Their technology can be performed silently to drill 1.5 cm holes at 2.45 GHz frequency. This apparatus has been successfully applied to a variety of materials such as glass, basalt, concrete, ceramics, and silicon, but it is still at the bench scale.

In conclusion, it can be inferred that although tools operate on different scales, their penetration and fragmentation functions occur through the application of energy on similar rock failure mechanisms. In any rock destruction method, processes conducting a very rapid application of energy are critical to produce failure. In drilling, cutting, breaking, kerfing, boring, or any other similar applications, the penetration rate "R" is directly proportional to the amount of energy "P" applied to the process. Therefore, microwave energy in combination with a breakage machine raises the total energy applied to the rock and results in an easier failure of the rock.

## 3. Review of recent research on microwave irradiation of rocks and minerals

The purpose of this section is to focus on various applications and to present a technical review of the state-of-the-art experimental studies, modeling and simulation literatures, as well as technoeconomic analyses introduced in the literature from 2000 to 2020. As a final contribution of this section, the challenges involved in this research area with some concluding remarks from the discussed literatures are debated. Moreover, the attempts that should be made to numerically model the process of microwave treatment of rocks via computer simulations for future predictions of the effects of microwaves on different rock types without costly (and sometimes impossible) experiments are discussed.

### 3.1. Experimental studies

In the 2000s, several studies highlighted the effects of microwave treatment on different types of rocks and minerals by performing various experimental tests. The experimental approach is an ongoing methodology used by researchers to verify the potentiality of microwave energy in the pre-conditioning and breakage processes of rocks and in the processing of minerals. In this section, after the primary experimental studies are extensively discussed and reviewed, a summarized list of the results published in secondary experimental studies on laboratory experiments of microwave treatment of rocks and minerals is given in Table 3. Because of the scarcity of the literature, the following review is organized so that each paragraph discusses all publications for each well-known research group or studies with the same perspective in the field.

From the beginning of the 2000s, researchers from the University of Nottingham, UK, carried out a number of related studies on the effect of microwave irradiation on mineral processing, including minerals separation, grinding, purification, and the reaction mechanisms of microwaves on various types of rocks and ores with different morphologies. Kingman et al. [71] investigated the grindability of ores according to the Bond Work Index (BWI). They experimentally irradiated four samples, including massive Norwegian ilmenite ore, massive sulphide, highly refractory gold ore, and open pit carbonatite, in a multi-mode microwave cavity with a variable power of 2.6 kW and a frequency of 2.45 GHz. From the results of microwave irradiation tests on the selected ore samples with an applied power level of 2.6 kW at different exposure times of 10s, 30s, 60s, 90s, 100s, and 120s, the authors concluded that microwaves can significantly reduce the BWI of ores, which implies that microwave treatment has the ability to decrease the energy required for fragmentation of ores. The same conclusion was later drawn in a study by Vorster et al. [25] on the grindability response of a massive copper ore. This study showed that the work index of massive copper ore may be reduced by up to 70% after 90s of microwave exposure at a power level of 2.6 kW in a multi-mode cavity. Later, Jones at al. [72] presented a review of microwave heating applications in environmental engineering and surveyed all related conclusions in studies prior to 2002. Jones at al.'s study concluded that an efficient design of microwave heating equipment is not possible without a proper knowledge of the material's dielectric properties. The authors further pointed out that a detailed fundamental knowledge of microwave engineering is needed for the development of an efficient design of microwave cavities to reduce the energy input for a pilot test or economically feasible industrial commercialization. Furthermore, in an investigation on the influence of microwave treatment on andesite specimens, Znamenácková et al. [73] exposed three individual cored samples. These samples were melted completely after being subjected to 1.35 kW at a frequency of 2.45 GHz after 10 minutes microwave exposure and 30 minutes with 2.7 kW at the same frequency in a multi-mode cavity. To investigate the results from this experiment, the same authors later performed a characterization method, an X-ray Diffraction (XRD) analysis, which revealed that the basic chemical composition of andesite remained unchanged; nevertheless, its structure became amorphous after the microwave experiments [24,74].

In 2004, Kingman et al. [75] experimentally studied the influence of short-term microwave treatments of copper carbonatite ores, up to 1 kg in weight, with power levels of 3 kW to 15 kW in a single-mode cavity. They conducted point load, drop weight, grindability, and liberation tests on the treated and untreated ores. According to the results of these experiments, the authors reported that by treating the samples for 0.2s exposure at 15 kW microwave power with an energy input of 0.83 kWh/t, 30% reduction in impact breakage parameters was achieved. Kingman et al. [75] further pointed out that by using a microwave-assisted comminution process, reduced plant size, potentially reduced wear costs per tonne, lowered water usage, and produced smaller downstream recovery circuits could be considered as benefits of microwave usage. In another study, Kingman and colleagues [75] experimentally tested the effects of microwave power on the breakage of lead-zinc ore samples by employing both single-mode and multi-mode microwaves with an operating frequency of 2.45 GHz at different power levels of 5, 10, and 15 kW for various exposure times of 1s, 5s, and 10s. The authors performed comparative tests on the results of their single-mode vs. multi-mode microwave irradiation tests and on their treated vs. untreated samples. From the results of multi-mode microwave tests, Kingman et al. [75] found that the strength of their samples was reduced rapidly at higher microwave power levels (e.g. 15 kW). However, multi-mode microwave irradiation tests with lower power levels were observed to be less effective. Subsequently, by using drop weight tests for quantification of the changes in the strength of the treated ore samples, up to 40% reduction in required comminution energy was achieved for ore particles of a mean size of 14.53 mm. Likewise, after single-mode microwave irradiation experiments at 10 kW, strength reduction of 50% was observed in the irradiated ores with a residence time of only 0.1s, which implies that for both single-mode and multi-mode microwave applications, a high input power level of microwave plays an important role for strength reduction and ore failure.

In 2005, experimental investigations on the effects of microwave treatment on the fracturing and grinding of kimberlite ore samples were studied by Didenko et al. [76]. They employed continuous wave magnetrons with power levels of 0.6 kW and 5 kW at a frequency of 2.45 GHz. The authors specifically mentioned that they used resonator-type rather than waveguide-type microwaves to elevate the microwave-energy density, which allowed them to reduce the heating time of samples. After exposure of kimberlite samples of a size 3-4 cm3 into the cylindrical H111 resonator for several seconds at a microwave power of 0.6 kW, the samples' temperature reached hundreds of degrees Celsius. Didenko et al. [76] mentioned that the heating process was done alongside a series of explosions, and the samples were split into 2-4 pieces. They further used a vapor bath method to investigate the effect of pressure rise due to the vaporization of water in the pores on the fracturing process; and found thermal shock fracturing could occur because of either linear expansion of solids in the heating process or the rapid evaporation of water in the rock pores.

Another study by Wang and Forssberg [77] presented microwave-assisted comminution and liberation tests on various minerals, including dolomite, limestone, copper ore, and quartz. Although it is not mentioned which types of microwave applicators were used for microwave irradiation tests in the article, it can be inferred that the authors employed a multi-mode microwave applicator for treatment of their samples. They employed a microwave system with a volume of 2.5 m3, a frequency of 2.45 GHz, and variable power levels of 3 kW and 7 kW for the following microwave exposures: 0, 5, 10, and 30 minutes. To investigate the influences of microwave treatments on the samples, dry ball milling tests for grindability evaluations and UCS tests for samples' strength measurements were performed. By comparing the results of the untreated with treated (7 kW/10 min) samples, the authors observed that the uniaxial compressive strength of quartz and limestone decreased from 50 MPa to 25 MPa and 40 MPa to 35 MPa, respectively. On the contrary, the UCS size of dolomite was found to be incremented. Then Wang and Forssberg [77] concluded that the coarser particles (−9.50 + 4.75 mm) of limestone and quartz were affected by microwave heating to varying degrees. In terms of liberation, the authors remarked from the grindability tests that the initiation of cracks by microwave heating would favor the reduction of energy consumption in comminution processes. In addition, a better and cleaner liberation of the mineral particles from the ore matrix can be obtained when a selective fracturing along the grain boundaries occurs. Overall, the study by Wang and Forssberg [77] indicated that minerals' particle size had a significant effect in both mineral pre-treatment by microwaves and comminution and liberation processes; therefore, the result was an increased fineness of the ground product. A similar investigation by Amankwah et al. [78] on improved grindability and gold liberation by microwave pre-treatment of a free-milling gold ore also verified the above findings.

In 2006, the influence of applying microwave irradiation with low power levels of 100 W to 150 W on basaltic rock was experimentally investigated by Satish et al. [79]. A multi-mode microwave cavity with an input power density of 1 W/g and an operating frequency of 2.45 GHz was used to irradiate cylindrical samples with a diameter of 38.1 mm and a height of 40 mm at exposure times of 60s, 120s, 180s, and 240s. The temperatures of the samples were recorded by using an Infrared (IR) camera after microwave treatments. Moreover, a standard point load tester was used for further mechanical and physical investigations on the amount of compressive and tensile strengths in the irradiated samples. The authors found that an increased exposure time resulted in increased temperatures, cracks, local spallation, and, subsequently, reduction of the rock's final strength. The trend of this reduction in the samples' compressive strength with respect to microwave exposure is shown in Fig. 4(a).

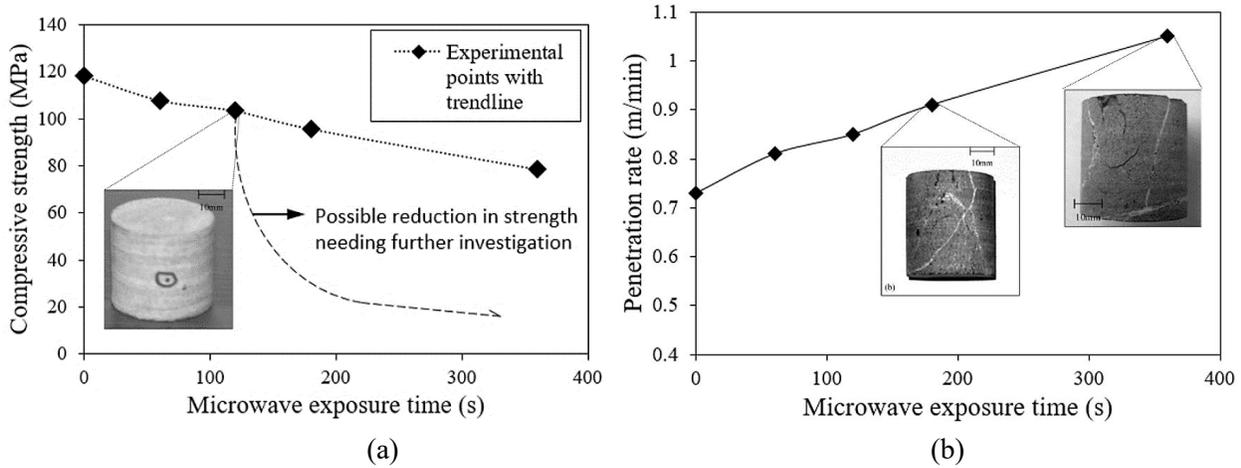

**Fig. 4.** Effect of microwave treatment of basaltic rocks with an input power density of 1 W/g at different microwave exposure times on (a) samples' mean compressive strength and (b) penetration rate of basalt for the percussive drilling process (modified from Satish et al. [79])

Satish et al.'s [79] study provides a plot of microwave exposure times in regard to penetration rate for the percussive drilling process. It has been shown that at 360s microwave exposure, a 42% increase in penetration resulted compared to nontreated samples. Furthermore, thermally induced cracks visually appeared only when samples were treated for 180s and 360s exposures (see Fig. 4(b)). In conclusion, this study verified that basaltic rock is very responsive to microwave exposures with low power levels as its temperatures linearly increase with time of irradiation.

Can and Bayraktar [80] studied the influences of microwave irradiation on the floatability and magnetic susceptibilities of different sulfide minerals including pyrite, chalcopyrite, galena and sphalerite. These selected minerals were irradiated in a multi-mode microwave system at power levels of 600 W, 950 W, and 1300 W and exposure times of 5s, 20s, 60s, 90s, 120s and 240s. The authors performed microfloatation and magnetic separation tests to investigate the effects of the applied microwaves on the minerals. From the results of the experiments, Can and Bayraktar concluded that the sulfide minerals were unequally affected by microwaves; the floatability of pyrite, chalcopyrite and galena was negatively affected by microwave treatment; the floatability of sphalerite became unchanged after microwave treatment; and finally, the magnetic susceptibilities of the sulfide and the oxide compounds formed on pyrite surfaces were higher than untreated pyrite. Later, a similar study by Waters et al. [81] also verified that thermal treatment has a considerable effect on the magnetic recovery of pyrite.

In 2009, the effect of mechanical property variation and the cutting rate on the microwave-treated granite samples were experimentally studied by Sikong and Bunsin [82]. They used a variable-power (up to 1000 W) multi-mode microwave oven with a frequency of 2.45 GHz for the treatment of orthorhombic shape granite samples with sizes of 15 mm × 50 mm × 70 mm at power levels of 600 W and 850 W and different exposure times. The results of this study show that the compressive strength of granite was reduced by 60% after 30 min at 850 W; the compressive strength of treated and quenched samples was reduced by 70% after 30 min at 850 W; and finally, the cutting rate of treated and quenched samples was reduced by 38% after 10 min microwave exposure at 600 W power. In a similar study, Peinsitt et al. [29] investigated the effect of microwave treatments on UCS, wave velocities, and heating characteristics of basalt, granite, and sandstone in their dry and water-saturated states using a multi-mode microwave oven with a power level of 3 kW and a frequency of 2.45 GHz. The researchers irradiated cylindrical shape samples, 5 cm in height and diameter, which were positioned in the center of the microwave cavity. After microwave treatments with different exposure times, the authors investigated the effects of microwave irradiation on their samples by measurements of temperature using an IR camera, visible modifications on the surface, UCS, and Ultrasound Velocity (USV). The results of surface temperature measurements are shown in Table 1. The main conclusion that should be drawn from the results of Peinsitt et al.'s [29] study is that severe damage (including large cracks and breakage) occurred in the samples when higher energy inputs produced high temperatures in the samples' interior. This conclusion shows that the generation of microwave-induced damage of rocks is mainly due to the large discrepancies between the temperatures of the interior and the outer surface of an irradiated rock.

**Table 1.** The results of highest surface temperatures achieved at recorded microwave exposure times in different rock samples after 3 kW microwave treatment [29]

| Rock sample | Sample's state | MW* exposure time (s) | Highest surface temperature achieved (°C) |
|---|---|---|---|
| Basalt | dried | 330 | 60 |
| | water-saturated | 325 | 60 |
| Granite | dried | 220 | 300 |
| | water-saturated | 295 | 300 |
| Sandstone | dried | 255 | 300 |
| | water-saturated | 125 | 30 |

* In this table, the word microwave is abbreviated as MW.

In 2010, Kobusheshe [83] experimentally investigated the effects of both single-mode and multi-mode microwave treatments on two kimberlite ores consisting of significant amounts of hydrated minerals. Point Load Strength (PLS) tests and Ultrasonic Pulse Velocity (UPV) measurements were used to evaluate the intensity of the damages induced within the samples because of microwave treatment. The author noticed high variability in the results and significant discrepancies between the mean and median values of the measured properties. It was noted that these variations were due to the anisotropic nature of the rocks and inconsistencies in electric field properties within the cavity. After different microwave power levels and exposure times were implemented, the results from the PLS tests were used to measure the strength of two kinds of copper and kimberlite ore samples. The results revealed that the stronger the ore samples, the higher the reduction. In addition, better results were achieved in the kimberlite samples that were treated for longer exposure times. The author concluded that the presence of hydrated minerals led to this phenomenon, which demonstrates the value of implementation of the right method of power delivery over higher power levels and microwave energy inputs. Furthermore, because of the incapacity of the point load test in observing the damage prior to and after the microwave experiments on the sample particle, UPV tests were conducted on the copper and kimberlite ore samples. As a result, Kobusheshe [83] noticed that a 10% reduction in the mean UPV in the kimberlite particles was promoted by using 6.81 kWh/t microwave energy input. In another study, the grindability of microwave treated iron ore samples was experimentally evaluated by Kumar et al. [84]. To this end, a variable power (up to 900 W) multi-mode microwave system with a frequency of 2.45 GHz was employed. After microwave treatment of iron ore at a constant power level of 900 W and various exposure times, the authors observed an increase in the ore's temperature. Consequently, the grindability of the iron ore was observed to increase significantly, with the specific rate of breakage rising by an average of 50%. Therefore, the authors pointed out that the grindability of microwave treated iron ores was accomplished much more rapidly than for untreated samples. In conclusion, it can be inferred from the results of Kumar et al.'s [84] study that microwave energy induces thermal stress cracks and, subsequently, decreases the energy required for grinding of iron ore.

In 2011, the potential application of microwaves in hard rock drilling, cutting, and breakage was introduced and analyzed by Hassani and Nekoovaght [17]. This study aimed at the development of microwave-assisted machineries to break hard rocks by investigating the effects of both single-mode and multi-mode microwave irradiation on temperature profiles and strength reduction (with UCS and BTS tests) in hard rocks for different power levels and exposure times. The authors inferred that only single-mode microwave applicators are able to provide high electric field intensity in order to break hard and abrasive rocks. Hassani and Nekoovaght's article indicates that by combining the microwave energy and one of the available excavation machines—Roadheader, Tunnel Boring Machine (TBM), or Oscillating Disc Cutter (ODC)—the total energy applied to the rocks can be decreased. The penetration rate consequently increases when an excavation machine is equipped with microwave technology. These results opened a new horizon for future breakage operations by one of the available rock excavation machines, i.e. drilling and/or full-face TBM, combined with rock pre-conditioning using microwave energy to perform a continuous non-explosive rock excavation. In another study, Rizmanoski [85] investigated microwave-assisted pre-treatment with modulated power levels on the breakage of copper ore. In this study, a single-mode microwave applicator with a frequency of 2.45 GHz was employed for microwave treatments of copper ore samples with different particle sizes. By performing comparative drop weight tests, the author found that the microwave-treated samples for 5s exposure at a modulated 5 kW power level break more easily than untreated samples. The objective of this study was to minimize the applied microwave energy using modulated power in order to obtain designated thermal stresses at a shorter exposure time. As a result, the modulated microwaves compared to continuous microwaves were observed to be energy savers in inducing the desired thermal stresses. Therefore, modulated microwave irradiation of ores can be effective at enhancing mineral processing. Three other studies [25,75,86] also verified this conclusion. In addition, a similar publication by Wang, Shi, and Manlapig [87] on pre-weakening of mineral ores by high voltage pulses also showed that pulsed waveforms can reduce energy consumption in the downstream grinding processes of ore minerals.

In 2012, Chen et al. [88] experimentally examined the behavior of ilmenite ores under microwave treatments with 3 kW power and a frequency of 2.45 GHz at exposure times of 10s, 20s, and 30s. After microwave irradiation tests, the irradiated samples were ground for 60s with a laboratory crusher. Chen et al. [88] characterized their treated ore samples by using the following methods: XRD, Scanning Electron Microscopy (SEM), and Fourier Transform Infrared (FT-IR) analysis. They found that, in general, ore minerals (e.g. basalt) responded more favorably to microwaves than gangue minerals. More significantly, their characterization results showed that the behavior of minerals with microwave irradiation was compositionally dependent. Therefore, the study demonstrated that the temperature dependency of the rocks' material properties plays an important role in the heating and pre-conditioning processes by microwaves. In another study, Hartlieb et al. [89] treated cylindrical samples of basalt experimentally in a multi-mode microwave cavity system to investigate the amount of heating and microwave-induced damage. After microwave treatments of basalt with 3.2 kW power at 10s, 20s, 30s, 40s, 60s, and 120s exposure times, the samples' surface temperatures reached 100 °C after 30s, 280 °C after 60s, and 450 °C after 120s of irradiation. Additionally, Hartlieb et al. [89] further indicated that the temperature in the samples' center reached 250 °C and 440 °C after 60s microwave exposure. From the results of the samples' heating and cracking behaviors, the authors concluded that the development of cracks in basalt is governed by macroscopic temperature gradients and the geometry of the sample instead of its mineralogical composition.

In another research project Jerby, Meir, and Faran [90] experimentally and theoretically demonstrated the thermal-runaway instability induced by localized microwaves in basalt. By employing a single-mode microwave cavity with a power level of 0.9 kW and a frequency of 2.45 GHz, a cubic sample of basalt was irradiated at different exposure times (up to 10 min). At 10 min microwave exposure, the sample's surface temperature was reached to slightly more than 1200 K, and the authors recorded the melting temperature value of basalt at approximately 1300 K. In another study, Swart and Mendonidis [91] evaluated the effect of radio-frequency pre-treatment of granite for comminution purposes. By comparing the results from SEM analysis of the untreated and treated granite samples, no significant changes in the form of fracture along the mineral grain boundaries were found in the treated samples in comparison to the untreated samples. Therefore, Swart and Mendonidis [91] concluded that electromagnetic waves within the Very High Frequency (VHF) range do not significantly weaken the mineral grain boundaries; and hence, the mineral liberation process is of no benefit.

In 2014, a review on the applications of microwave energy in cement and concrete was presented by Makul, Rattanadecho, and Agrawal [92]. Similarly, another review of microwave processing of materials and applications in manufacturing industries was presented by Singh et al. [93]. In another publication, Irannajad et al. [94] experimentally tested the floatability of ilmenite ores after microwave treatments in a multi-mode microwave cavity with different power levels (up to 1000 W) and an operating frequency of 2.45 GHz. To optimize microwave irradiation time of the ores, several experiments for different exposure times (up to 600s) at the highest power level of 1000 W were carried out. The authors observed that the flotation recovery of ilmenite was improved by increasing the microwave exposure and a maximum recovery of ilmenite (94%) after irradiation for 150 seconds was obtained. Longer exposures had no significant impact on ilmenite flotation recovery. However, it should be noted that when considering the potential for microwave-enhanced liberation of flotation recovery, several other factors, such as the mineral's grain size, grind size and valuable mineral associations, must be well understood [95,96]. In other experimental work, Like, Jun, and Pengfei [97] exposed a UCS size sample of granite to microwave treatments with a power density of $8 \times 10^6$ W/m$^3$ at different microwave irradiation times. Following the results of the experiments, a linear decrease in rock mass strength was achieved with increasing microwave irradiation time and power density. From microwave irradiation time changing from 0s to 60s, the mass strength of granite was decreased from 50.6 MPa to 29.6 MPa, which is 41.5% in strength reduction. Thus, the researchers concluded that microwave irradiation effectively reduces rock mass strength; therefore, the longer the microwave exposure time, the lower rock mass strength.

In 2016, a technical research study investigating the effects of different microwave power levels and exposure times on temperature profiles and strength reduction in hard rocks was presented by Hassani et al. [4]. They treated cylindrical and disc-shaped samples from three different rock types (two types of basalt from different places, mafic norite, and granite) under microwave treatments with power levels of 1.2, 3, and 5 kW for 10s, 65s, and 120s in a multi-mode microwave cavity. They used cylindrical disks, 50 mm in diameter, to perform strength tests on their samples. The disks were 100 mm high for UCS tests, and shorter disks at the height of 25 mm were used for BTS tests. According to their work, the maximum energy of 740 kWh/t was added to the disks during the microwave pre-treatment. The authors observed that low exposure times of about 10s did not lead to any significant impact on the tensile strength of the samples.

Microwave treatment did not affect the UCS of the granite samples, and the basalt samples showed a 30% decrease under 65s irradiation with an input power of 5 kW. The results of the BTS tests showed that increasing the exposure time had the same effect on strength reduction, although the energy input to the samples was increased. Moreover, microwave treatment had a better effect on norite than granite and basalt in terms of strength reduction. Hassani et al. further conducted single-mode microwave experiments—in addition to multi-mode microwave experiments—to measure the surface temperature distribution on a rock positioned at different distances from microwave horn antenna. They performed several tests on slab-shaped basalt specimens (40 × 40 × 40 cm$^3$) at a constant 3 kW power for 60s and 120s exposures and at six distances of 3.5, 6.5, 9, 12, 15, and 19.5 cm. The local damage spread approximately 3 cm in depth, whereas the global damage exceeded 10 cm. Hassani et al.'s [4] study shows that exposure of rocks to high-power microwaves results in increased temperature and formation of mechanical damages to the rock surface, better facilitation of breakage operations for the mechanical disc cutter of an excavation machine, and reduction of disc cutter wear. According to the results presented by Hassani et al. [4], a combination of microwaves with one of the available breakage machines, such as the TBM, could improve the rate of penetration, thereby increasing the lifetime of the cutter and saving on the overall cost of the project.

Another research project by Hartlieb, Toifl, Kuchar, Meisels, and Antretter [30] experimentally addressed the temperature dependency of the electrical, thermal, and mechanical properties of basalt, granite, and sandstone. The authors measured thermo-physical properties of their selected rocks, such as specific heat capacity, thermal diffusivity, thermal expansion, thermal conductivity, and dielectric properties, in the temperature range of 25 °C to 1000 °C, thus showing how phase transitions influence the rock's texture and stability. According to Hartlieb et al.'s [30] measurements, the strong variation of microwave effects depends mainly on the rocks' dielectric properties.

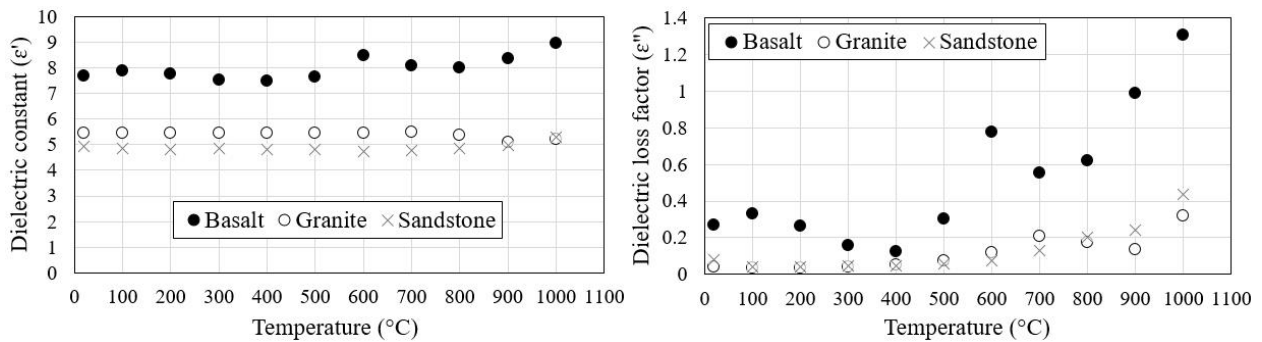

**Fig. 5.** Temperature dependency of dielectric constant (left) and dielectric loss factor (right) of basalt, granite, and sandstone in the temperature range of 25-1000 °C at a constant frequency of 2.45 GHz (data taken from table in Hartlieb et al. [30])

To better illustrate how dielectric properties of the selected rocks varied at different temperatures, Hartlieb et al.'s [30] tabular findings are plotted in Fig. 5. As shown, both plots of the dielectric constant and the loss factor of basalt achieved higher values than granite and sandstone. Moreover, a similar behavior can be observed from the plots of granite and sandstone samples, which means that the responses of these two rocks to microwave treatment should not vary significantly temperature wise. Overall, the data of Hartlieb et al.'s [89] study shows the impact of the rock forming minerals on the structure and thermal behavior of selected hard rocks; and therefore, selective heating by microwave irradiation is possible when rocks are made up of different minerals (absorbents) with different dielectric properties.

In 2017, a study by Hartlieb, Grafe, Shepel, Malovyk, and Akbari [98] addressed the crack patterns caused by microwave irradiation and the effect of these cracks on subsequent breakage. Their results included crack pattern and damage propagation, cutting force distribution, specific energy consumption, and particle size distribution for nontreated and treated granite samples for 30s and 45s. Hartlieb et al. [99] used $50 \times 50 \times 30$ cm$^3$ sized Neuhauser granite and exerted a UCS of 210 MPa and Cerchar Abrasivity Index (CAI) of 4.2. The authors reported that the local damage spread approximately 3 cm in depth, whereas the global damage exceeded 10 cm. They also noted a 22.5% reduction in average cutting forces due to the microwave treatment of the sample for 45s when the cutting forces were reduced from 6.04 kN and 6.26 kN to 4.26 kN and 5.22 kN for 8 mm and 12 mm cutting spacing, respectively. Their experiments included treatment of 18 spots on the granite samples with a 24 kW microwave system operating at 2.45 GHz for 30s and 45s. Their results showed a 4.7 kWh/t conservation of cutting energy achieved because of the reduction in cutting force, in the best case, at a treatment time of 45s. Hartlieb et al. [99] also used a 24 kW microwave with a frequency of 2.45 GHz to create a network of cracks on the surface of granite with a UCS of 210 MPa and a CAI of 4.2. The authors demonstrated that the local damage (created around the microwave treated area) and global damage (generated in the area next to the treated surface) after 45s of irradiation decreased the cutting forces and drilling energy consumption dramatically, equaling to a greater penetration rate. Other studies [10,99] have also verified this conclusion. In addition, a comparative analysis on the effect of microwave irradiation on the milling and liberation characteristics of minerals with different morphologies was carried out by Singh et al. [100]. The study shows how different microwave energy inputs affect minerals and their processing operations. To this end, three different types of minerals, including a coal sample, iron ore, and manganese ore, were treated in a multi-mode microwave cavity with an operating frequency of 2.45 GHz at different power levels of 180, 540, and 900 W and exposure times of 1, 3, and 5 minutes. After irradiation tests, the samples' overall temperatures were recorded by a thermocouple and reported by the authors. However, for a better understanding of how the different samples in Singh et al.'s [100] study responded to microwave treatments with different energy inputs and exposure times, the results in their table of temperature data are plotted in Fig. 6 for different microwave testing conditions.

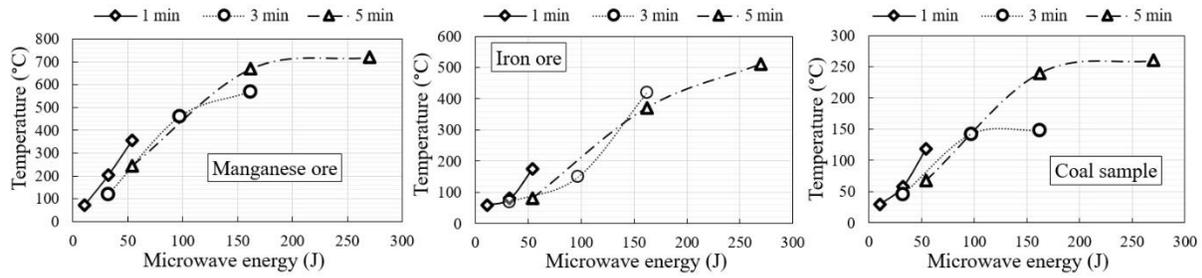

**Fig. 6.** Effect of microwave treatments with different energy inputs and exposure times on the samples (data taken from table in Singh et al. [100])

By comparing the plots, it can be seen that the temperatures of all three samples increase with increasing microwave energy inputs. However, with the same amount of microwave energy input, manganese ore heated more than the other two samples. The best operating ranges of input microwave energy for treatment of coal, iron ore, and manganese ore were found to be 1 Wh to 30 Wh, 5 Wh to 45 Wh, and 15 Wh to 40 Wh, respectively. The authors concluded from the results of the sample characterizations that microwave treatment improved the grinding of coal, manganese ore, and iron ore. Moreover, liberation studies demonstrated that microwave treatment can lead to a 17.1% increase in carbon recovery for coal and a 42.46% increase in manganese recovery, but the iron ore did not show any improvement in mineral liberation.

In addition, a possible design of several single-mode microwave antennas on breakage machines such as cutting discs and TBMs for microwave-assisted excavation was developed and proposed by Zheng [101]. The study demonstrates that it is feasible to install several microwave applicators on a cutterhead. However, there might be a challenging problem for a rotating cutterhead because of the very short exposure time of microwaves, which reduces the amount of heating and the subsequent micro fracturing effects. Another significant result from Zheng's study is his recommendation to use a horn applicator in order to have better directionality when microwave applicators (antennas) are installed on TBM cutterheads. In an extended work, Zheng et al. [102] experimentally investigated the effect of microwave treatment on thermal and ultrasonic properties of gabbro using a 2 kW single-mode microwave at a frequency of 2.45 GHz. Zheng et al.'s [102] study shows the effects of microwave irradiation with variable power levels of 0.5-2 kW and exposure times of 30-120s on the temperature, microcracks, and p-wave velocity of the samples using an infrared camera, an X-ray microscope, and an ultrasonic pulse transmitter. In terms of microcracking analysis of the irradiated specimens, the evolution of cracks at 2 kW power started as the microwave duration increased from 30s to 120s. The authors also stated that they observed both intergranular and transgranular cracks; however, intergranular cracks were more dominant. Another significant conclusion was that the density of both macro- and micro-cracks increased with increasing microwave exposure, which indicates that a considerable reduction in the mechanical strength of the rock was achieved. From the data of p-wave velocity tests, Zheng et al. [102] observed that the overall p-wave velocity in their samples was reduced by up to 55%, which implies a significant reduction in the strength of the samples. Finally, it is concluded that a pulsed microwave with a high power level at a short exposure time yields better rock heating and pre-conditioning, and this can be considered as an alternative method of rock breakage with continuous microwave irradiation.

In 2018, in an extended research project, Hartlieb and Rostami [10] presented initial results of their laboratory-scale microwave irradiation experiments on a block of granite using a high power (24 kW) microwave apparatus. The objective of this study was to present innovative concepts for improving the breakage of hard and abrasive rock types by pre-conditioning and inducing micro-cracks with high power microwaves. The authors introduced the Rock Mass Rating (RMR) system to quantitatively express the condition of the irradiated rock mass related to its strength and mechanical behavior. From the results of applying high power microwave irradiation on granite, the authors concluded that (1) an extended network of cracks occurs in hard rocks such as granite, basalt, quartzite, etc., because of high power irradiation, (2) induction of micro and large cracks reduces RMR values and can influence both rolling and normal forces of the cutting discs, and (3) the reduced cutting forces and power consumption lead to an increased penetration rate of the breakage machine, which ultimately increases cutter life. These findings are good indicators for evaluating the feasibility and possibility of applying high power microwaves on future breaking machines such as TBMs. In another project, Hartlieb, Kuchar, Mosar, Kargl, and Restner [103] performed low power (3.2 kW) multi-mode microwave irradiation tests on various types of rocks (hard and abrasive) to assess their physical and chemical changes before and after microwave treatments with different exposure times. To this end, cylindrical rock samples with the same size (d = 50 mm, h = 50 mm) were used. The selection of the given rocks for microwave irradiation experiments was based on different considerations i.e. rocks in terms of hardness and abrasivity and the rocks' ability to absorb microwaves. In this case, a quick overview of Hartlieb et al.'s [103] categorization of rocks is given in Table 2. After microwave irradiation experiments on the selected rock types, the authors conducted XRD analyses, p-wave velocity measurements, and surface temperature and ultrasound velocity measurements from the treated rock samples, which resulted in the following findings. Samples of basalt heated up from their initial temperature and, therefore, microscopical damage (crack) occurred after 40s of microwave exposure. Extensive and wide cracking was observed in the treated samples after 120s of irradiation; and therefore, ultrasound velocity of the samples significantly decreased. Similarly, an increase in average surface temperature of the irradiated gabbro sample was observed with increasing irradiation times. This surface temperature rise was accompanied by a decrease in the gabbro's p-wave velocity.

**Table 2.** Categorization of different rock types from data in Hartlieb et al. [103]

| Rock name | Rock mineralogical type/texture | Microwave absorbability | Dielectric constant, ε' | Dielectric loss factor, ε" |
|---|---|---|---|---|
| Granite | Mafic volcanic, coarse-grained texture | low absorber | 5.0-5.8 | 0.03-2 |
| Sandstone | Mafic volcanic, medium-grained texture | low absorber | 4.93 | 0.08 |
| Copper ore | Hard and tough | good absorbed under specific circumstances | High | High |
| Basalt | Greenstones, fine-grained texture | relatively good absorbers | 5.4-9.4 | 0.08-0.88 |
| Diabase | Greenstones, fine- to medium-grained texture | relatively good absorbers | 5.4-9.4 | 0.08-0.88 |
| Gabbro | Medium-grained texture | good absorber | 5.4-9.4 | 0.08-0.88 |

Hartlieb at al. [103] concluded that rocks with similar absorption properties are strongly influenced by rock texture. They further pointed out that coarse grained rocks (i.e. gabbro) showed cracks along grain boundaries, but random crack networks were found to be in the fine-grained rocks, such as basalt. Similarly, the grain size of each rock type played an important role in the rock's heating, as highly absorbing copper ore was not influenced by microwaves in comparison to the fine-grained and readily heated ore. Further analyses and investigations on quantification of cracks in rocks after microwave treatment and a comprehensive review of different techniques for characterizing cracks induced in rocks with can be found in a review study by Nicco et al. [104]. Another study by Forster, Maham, and Bobicki [37] experimentally performed microwave heating tests and high-temperature dielectric property (real and imaginary permittivity) analysis on magnesium silicate minerals. The findings showed how microwave irradiation affects the microwave heating properties of serpentine and olivine (the primary components of ultramafic nickel ores and constituent minerals). The authors experimentally irradiated cylindrical samples of serpentine and olivine (with a diameter of 16 mm, a mass of 7.5 g, and heights of 17.5 mm and 15 mm, respectively) in a multi-mode microwave cavity with a power level of 1200 W and a frequency of 2.45 GHz. Then, by using the cavity perturbation technique, they measured samples' dielectric properties, involving real and imaginary primitivities at different frequencies and different temperatures. The dielectric properties of olivine and serpentine were found to be frequency dependent, especially in the serpentine curves at low temperatures (below 500 °C). In addition, frequency dependence was also observed for the imaginary permittivity of serpentine at temperatures above 800 °C. Comparatively, the imaginary permittivity values for serpentine were significantly higher than those for olivine across the temperature range 0-1200 °C (see Fig. 7). The results of Forster et al.'s [37] study demonstrate that for serpentine, the imaginary permittivity (ε") —or simply the loss factor—largely decreased with increasing temperature in the microwave heating tests, resulting in a decreased heating rate. For olivine, the imaginary permittivity was nearly zero in the range of temperatures; thus, the heating rate was also close to zero.

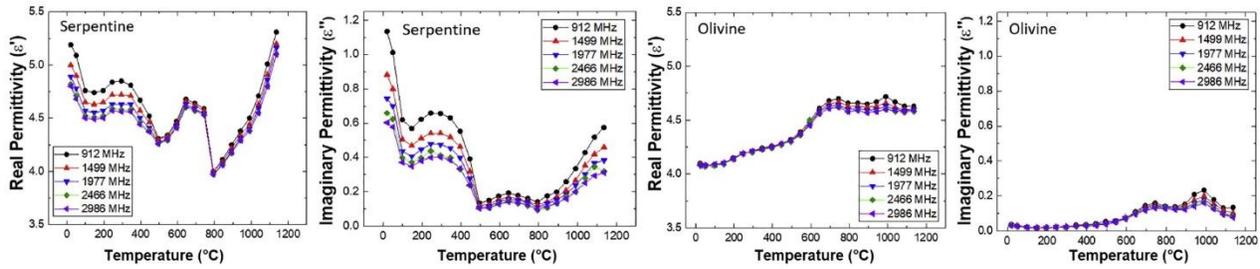

**Fig. 7.** Real and imaginary permittivity values of serpentine (left) and olivine (right) as a function of increasing temperature [37]

A similar work by Bobicki, Liu, and Xu [105] focused on the effects of microwave treatment of two types of ultramafic nickel ores: Okanogan nickel (OK) ore and Pipe ore—obtained from the Vale-owned Pipe deposit—on the overall process improvements, including grindability, rheology, flotation, material handling, dewatering and tailings treatment. To this end, the authors employed a 1000 W multi-mode microwave with a frequency of 2.45 GHz. The ore samples were then treated with 100% microwave power capacity. The temperature results of the OK and Pipe ores with respect to microwave heating time are illustrated in Fig. 7(a). The plot indicates that the OK ore sample has a high initial heating rate that declined with increasing temperature and microwave exposure time. The same trend can be observed from the plot of the Pipe ore up to 8 min microwave heating time. However, it is important to note that the authors mentioned that after 8 minutes of exposure, a partial melting was observed in the Pipe ore sample. This occurrence resulted in a change in the trend. To investigate why there were differences between the obtained heating rates in the two ores, Bobicki et al. [105] further experimentally measured imaginary and real permittivities of the ores with increasing temperatures at frequencies of 912 and 2466 MHz, as shown in Fig. 8(b) and 8(c). It can be seen that both imaginary and real permittivity of the ores vary with increasing temperature. However, the measured values for Pipe ore at the frequencies of 912 MHz and 2466 MHz are higher than the OK ore at temperatures between 0-400 °C and 800-1100°C, respectively. Therefore, as power dissipation in rocks is highly influenced by the rock dielectric properties, the obtained variation in imaginary and real permittivity of the two ores is the main reason for the difference in their heating rates. Further analyses on the treated ore samples were conducted by Bobicki et al. [105] using the following techniques: X-ray Fluorescence (XRF) spectroscopy, quantitative XRD analysis, Fourier Transform Infrared (FTIR) spectroscopy, and SEM methods. From the results of ore characterizations after microwave treatment, the authors concluded that by microwave irradiation: (1) the grindability of ore with consistent texture (OK ore) improved; (2) the grindability of ore with inconsistent texture (Pipe ore) decreased; and (3) the specific surface area of both ores improved. Ultimately, the authors reported that since microwave pretreatment did not decrease the energy required for grinding under the specified microwave conditions, the energy savings might only be realized for overall process improvements (e.g. grindability, rheology, and flotation).

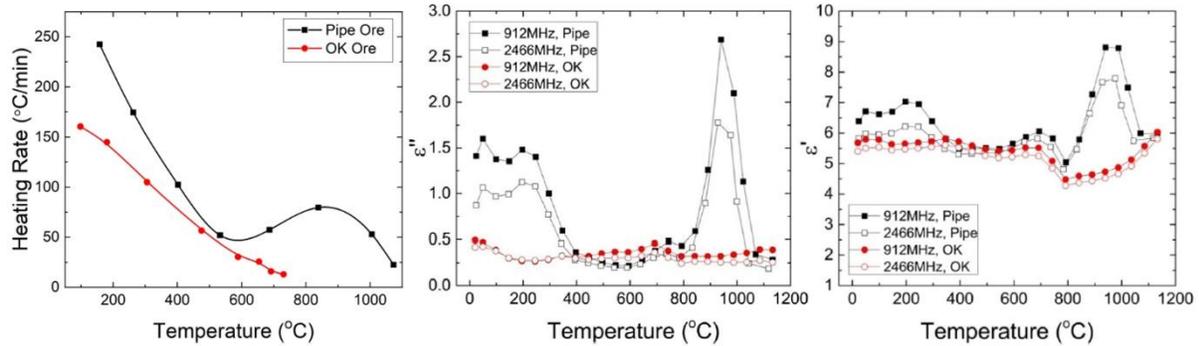

**Fig. 8.** (a) Heating rate upon exposure to microwave radiation with respect to temperature for Pipe (■,□) and OK (●,○) ores; (b) imaginary permittivity, and (c) real permittivity for the Pipe and OK ores with increasing temperature at frequencies of 912 and 2466 MHz [105]

Ong and Akbarnezhad [106] experimentally investigated various effects of microwave irradiation on samples of concrete. According to the results of their study, the variations in concrete's dielectric constant and its loss factor for a typical specimen with water content and microwave frequency were observed. Both dielectric constant and loss factors of the concrete increased significantly with an increase in the microwave frequency or water content, leading to a higher heating potential when exposed to microwaves. For a sample of concrete, there was an inverse relationship between the penetration depth of microwaves in the concrete and microwave frequency. This means that the higher the microwave frequency, the less the penetration depth. Therefore, at higher microwave frequencies, the microwave energy is expected to be dissipated within the surface layer of the concrete rather than penetrating deeply into the specimen; the higher the microwave frequencies, the thinner the affected surface layer. A review of the applications of microwave energy in rock and concrete processing is given by Wei et al. [107].

In 2019, structural changes in samples of granite (cube of 20 mm) at high temperatures (300-800 °C) induced by microwave irradiation were studied [108]. In this case, an adjustable multi-mode microwave (1.4 kW) with a frequency of 2.45 GHz was used for irradiation tests. The microwave treated samples were then characterized by SEM analysis, Thermogravimetric Analysis (TGA), Differential-Scanning Calorimetry (DSC) tests, X-ray Powder Diffraction (XRPD) analysis, and UCS tests. Thus, from the results of these rock characterization analyses, Zeng et al. [108] concluded that (1) granite samples were observed melted and cracked at 600 °C and completely melted at 800 °C, (2) both transgranular and dominant intergranular cracking modes were observed by the SEM, (3) the results from TG-DSC tests indicated moisture releasing, (4) fieldspar and biotite melted at 800 °C according to the XRPD results, and finally (5) the uniaxial compressive strength of granite decreased from 88.17 MPa at 25 °C to 18.61 MPa at 800 °C. The results presented in a similar publication by Li et al. [38] also arrived at the same conclusions. Additionally, Lu, Feng, Li, Hassani, and Zhang [109] developed a research study to examine experimentally the effects of different microwave power levels on the mechanical strength, burst time, and fragmentation of compact basalt samples comprising plagioclase, enstatite, olivine, and a small amount of ilmenite. A continuous wave multi-mode microwave cavity with a power range of 1 kW to 6 kW and the frequency of 2.45 GHz was used to treat the basalt samples at different power levels and exposure times. Cylindrical samples (50 × 100 mm) for the UCS, discs (50 × 50 mm) for the BTS and the PLS tests, and cube samples with four different length sizes of 30, 50, 75, and 100 mm for fragmentation tests were used. The microwave treated specimens were cooled to room temperature and placed for the tests in a loading direction perpendicular to the main crack direction in the BTS and PLS tests. A faster reduction in strength was observed at the higher microwave power levels. The authors concluded that lower energy consumption was needed to burst the specimens at higher power levels. Overall, Lu et al.'s [109] work reveals that all the measured values of UCS, BTS, and PLS tests on basalt were decreased by increasing the microwave irradiation time. In an extended study, Lu, Feng, Li, and Zhang [110] examined the microwave-induced fracturing of hard rocks for underground engineering applications by developing a novel (open-type) microwave-induced fracturing apparatus with an operating frequency of 2.45 GHz and variable power levels of up to 15 kW. Their microwave apparatus comprised single-mode microwave applicators to investigate the subsurface and borehole fracturing effects on basalt samples. The researchers performed both laboratory-scaled microwave experiments and field tests. Therefore, the uniqueness of this study was its field tests of microwaves at 15 kW power level on rock masses. From the results of laboratory-scaled microwave experiments, the authors concluded that the longer microwave exposures resulted in greater reduction in p-wave velocity. From the results of the field tests, the borehole fracturing mode rendered a favorable fracturing effect on boreholes. Moreover, a reduction in the sound velocity around the borehole and between the boreholes was observed. Overall, Lu et al.'s [110] study demonstrated a real application of microwaves in the field, which had never before been implemented to this extent. In the studies mentioned above, microwave-induced fracturing of basalt was successfully surveyed and analyzed. However, the only lack in the two studies might be energy and economical analyses for industrial implementations.

In 2020, several researchers and research groups have investigated the impact of microwave treatment on hard rock pre-conditioning and fracturing. For example, by using the same microwave system explained in the study by Lu et al. [110], the response of compact basalt samples under different confining pressures was surveyed by Lu, Feng, Li, and Zhang [35]. The study enhances the understanding of microcracking behavior of basalt after microwave treatments with different power levels (1 kW to 5 kW) and exposure times (up to 300s). Kahraman et al. [111] studied the effect of microwave treatment on the compressive and tensile strength of nine different igneous rock samples consisting of six granites, two syenites, and one gabbro. UCS and BTS tests on these samples before and after microwave treatments with power levels of 1 kW, 2 kW, and 6 kW at different exposure times ranging from 60s to 420s were performed. Kahraman et al. [111] observed that the UCS values of a granite sample, including a small but significant amount of microwave absorber minerals, decreased at surface temperatures above 200 °C. However, the BTS test values started to decline at temperatures above 100 °C for the same granite sample. The authors observed inconsistency and fluctuations in average temperature plots of the treated samples in their analysis of the experiments using an infrared gun after the microwave treatment. They concluded that depending on the mineral contents of each rock type, the heating degrees of the rocks differ from one type to another.

As surveyed in this section, many studies have highlighted the existing challenges and potential applications of microwaves in the rock breakage and mineral processing industries in order to increase mineral compound liberation as well as potential reduction of energy usage. Below, a technical summary, including the details of applied experimental methods and the main findings of secondary experimental studies not discussed in this section, but that have made significant contributions in the application of microwaves in rock mechanics and mining industries, is given in Table 3.

**Table 3.** Secondary publications found through a chronological literature review on laboratory experiments of microwave treatment of rocks and minerals and the corresponding effects

| Microwave operating parameters and sample info | Subject of study | Experiment methods | Main findings and highlights |
|---|---|---|---|
| MW treatment of small coal samples (19.05–12.7 mm) with a power level of 900 W; **M** | The influence of microwave pre-treatment on the grindability of coal | Sahoo et al. [112] Grindability tests, SEM, XRD | The MW treated coal was grinded more rapidly than untreated coal. With the specific rate of breakage increased by an average of 15% after MW heating, the grindability of coal increased significantly. |
| MW treatment of a porphyry copper ore with different sizes; with a power level of 15 kW (approximately 2 kWh/t); **S** | Study on how increasing grind size effectively influences liberation and flotation of the ore by MW treatment | Batchelor et al. [95] MLA, PLS test, crushing and grinding | An indirect reduction in specific comminution energy by grinding the ore for a shorter time might occur as a result of increase in grind size of the ore. |
| MW treatment of granite block samples with different sizes with a power level of 24 kW and 2.45 GHz frequency; **S** | Investigating the strength reduction of the rock by artificially induced crack patterns via MW pre-conditioning | Hartlieb et al. [99] X-ray CT, UCS, CAI, linear cutting test rig | After exposure of the block sample to 24 kW and 45s MW irradiation, the mean cutting forces were reduced as follows: from 6.04 and 6.23 kN down to 4.26 and 5.22 kN. |
| MW treatments of coal core samples (50 mm diameter and 60 mm height) with various power levels (2-10 kW); **M** | Evolution of pore structure under MW heating for coal with different water saturation conditions (from 1% to 15%) | Li et al. [113] NMR, X-ray CT, p-wave | The study shows the total porosity of coal rose linearly with MW power, while it increased exponentially with greater water content (1% to 15%). |
| MW treatment of granite cube samples (20×20×20 mm$^3$) with an adjustable power level of 1.4 kW and 2.45 GHz frequency; **M** | Determining the pre- and post-microwave microcracking and strength reduction of the samples at high temperatures induced by MW treatments | Zeng et al. [108] XRF, SEM, TGA, DSC | Both transgranular and intergranular cracking modes were observed from SEM images, but the latter dominated. The moisture releasing and α-β quartz transition was detected from TG-DSC result. |
| MW treatment of various ore particles at different sizes with a power level of 1.2 kW for 12s exposure and 2.45 GHz frequency; **M** | Investigating the position- and cross-dependencies of the rock particles in MW cavity for maximum heating; and the effect of particle properties (particle size, magnetic properties, thermal response) on the heating of adjacent particles | Jokovic et al. [114] MLA analysis, IR imaging | "If two highly MW responsive particles were in electrical or magnetic contact, they would be much more effective at absorbing energy than either particle on its own and cause even larger distortions in the applied MW field" (p. 6). |
| MW treatment of three igneous rocks (gabbro, monzonite, and granite) (W84×L41×H30 mm$^3$) with various power levels (0.5-2 kW) for various exposures (30, 60, 90, and 120s) at 2.45 GHz; **S** | Investigating the thermal, mechanical, and cracking behavior of igneous rocks after MW treatments with different power levels and exposure times | Zheng et al. [115] Ultrasonic s-wave and p-wave velocity measurements, UCS, UV, IR, and PPL imaging | Gabbro and monzonite specimens were thermally cracked and melted at MW treatment with 2 kW for 120s. Heating at higher power levels for shorter durations resulted in a more weakening. |

*Note: In the table the term "microwave" is abbreviated as MW; and the two MW applicator types of single-mode and multi-mode are denoted by **S** and **M**, respectively.

### 3.2. Modeling and simulation literatures

Computers have evolved and advanced significantly over the decades since they originated. As a result, numerical modeling techniques have become stronger in recent years for simulation of the process of microwave treatment of materials. In tandem with experimental research, computational studies have come to play an important role in understanding how different mechanisms are involved in microwave-assisted rock pre-conditioning and breakage processes. Numerical models can be used to predict the coupled Electromagnetic, Thermal, and Mechanical (ETM) multiphysics interactions in rocks under microwave treatments with different operating parameters including power levels, exposure times, and distances from the antenna (for single-mode microwaves). Moreover, the numerical modeling approach has been considered a useful comparative tool that allows quantification of the relationships between the material properties of rocks made up of different absorbent minerals and microwaves with different operating parameters. Therefore, the development of numerical models is considered an enhancement in understanding microwave-assisted rock breakage systems and their thermomechanical characteristics, i.e. temperature changes and initiation of mechanical stresses. In the 2000s, several studies have tried to numerically simulate the process of microwave heating and pre-conditioning of rocks and minerals to predict microwave effects. These modeling studies and their analyses for solving applied problems are now essential for a proper design and practical implementation of a functional microwave-assisted rock pre-conditioning system. The following literature review shows how evolution of computers and simulation techniques have significantly improved numerical modeling of microwave irradiation of rocks over the past two decades.

In 2002, a very simple numerical simulation of the microwave-assisted drilling process in alumina was developed by Grosglik, Dikhtyar, and Jerby [116]. They employed the FDTD method for their modeling in MATLAB software. However, because of the lack of computation capacity at the time of this work, only temperature and electric field at a concentrated point were shown. Later in 2003, by using the FLAC 2D commercial software, Whittles et al. [23] numerically modeled 2D coupled electromagnetic, thermal, and mechanical multiphysics to predict the influence of microwave power density (2.6 kW and 15 kW) and exposure time on the strength change of rocks after microwave treatment using the finite difference method. Their study used a "theoretical" ore consisting of a microwave absorbing pyrite mineral in a low-absorbing calcite matrix and considered the following two different numerical scenarios applied at the same frequency of 2.45 GHz. In the first scenario, it was assumed that samples of pyrite were already exposed to multi-mode microwave treatments at 1s, 5s, 10s, 15s, and 30s and a varied power density from $3 \times 10^9$ W/m$^3$ at 300 K to $9 \times 10^9$ W/m$^3$ at temperatures greater than 600 °C. Secondly, to study the effect of higher power levels of microwaves on temperature distribution, uniaxial compressive strength, and shear plane development within the ore samples, the study assumed that the samples were already exposed to single-mode microwave irradiation with a power density of $1 \times 10^{11}$ W/m$^3$ (15 kW at 2.45 GHz). Because of the higher power density, Whittles et al. [23] found that the strength of their simulated rock model dropped as the time of irradiation increased. The results presented in the Whittles at al.'s numerical study demonstrate that higher power density generates a considerably larger reduction in strength with much lower energy inputs, thus reducing energy requirements. Satish [117] built upon the work of Whittles et al. [23] by modifying the source of irradiation from a power density to an energy density to show the generation of stress within the grain boundaries of a single pyrite hosted calcite with the same material that had been defined by Whittles et al. [23]. The results of Satish's [117] work showed that a large amount of potential stress could be generated at the particle boundaries of pyrite because of the transparency difference between the two minerals. In addition, Jones et al. [11] developed a numerical model to study the effects of microwave power density and particle size on microwave treatment of ores prior to grinding. In contrast to the normal constant microwave energy previously used by Whittles et al. [23], the influence of pulsed microwave energy at five different irradiation times from 0.1s to 10s was employed by Jones et al. [11]. The results of their study illustrate that (1) the obtained stresses along the mineral boundary were predominantly shear and tensile in nature; (2) by increasing microwave power density, minerals absorb more heat, resulting in better mineral liberation due to the creation of more thermal stress through the minerals' particles; and (3) by varying mineral particle size, the overall fracturing process was influenced. Jones et al.'s [11] study concluded that a decrease in mineral size and an increase in input microwave energy are required for sufficient fracturing. In a later publication, Jones, Kingman, Whittles, and Lowndes [22] studied numerically and extensively the effects of microwave power density and exposure time on the mineral weakening process by microwaves. By using FLAC 2D software, the researchers developed 2D simulation of a simplified two-phase mineral ore comprising pyrite particles randomly disseminated in a matrix of calcite. The models were undertaken with both continuous and pulsed wave simulations. The authors simulated six power densities, varying between $1 \times 10^9$ and $1 \times 10^{10}$ W/m$^3$ for heating times varying between 0.1s and 10s in their continuous wave simulations. For pulsed wave simulations, the power density was varied between $1 \times 10^{13}$ and $2 \times 10^{15}$ W/m$^3$ with pulsed duration between 0.1s and 10μs. According to the results of Jones et al.'s [11] study, with pulsed microwave, lower temperature was required to achieve the same reduction in a sample's strength and was observed as a result of continuous wave simulation. Furthermore, with increasing power density, mechanical stresses increased, which resulted in a greater damage within the ore and, in turn, a lower UCS.

In 2006, to test the potential application of multiple ports in traveling waveguide applicators, Balbastre et al. [118] presented a simple numerical simulation of microwave heating in a cylindrical sample (5 cm diameter and 1 cm height) made up of a high loss material. By using the commercial ANSYS software and employing the finite element method, the authors were able to obtain the electric field amplitude in a closed region. Then, they found that a variable feed, such as ore particles in the design of microwave applicators, can be optimized for a specific load property, which enhances physical processes, such as the separation of minerals. The authors concluded that because of the limited electrical field uniformity within the material under microwave heating, the heating of high loss materials is a complex task. Typically, the limited electrical field uniformity is observed in large samples of low thermally conductive materials, as the homogeneity of microwave heating will be completely controlled by electrical field distribution, which increases the chance of hot-spots development or thermal runaway [119]. Ali and Bradshaw [120] numerically investigated the effect of ore texture on the amount of damage in a microwave treated ore model. By using FLAC 2D commercial software and employing the finite difference method, two binary ore models made up of galena-calcite and magnetite-dolomite were constructed. The models were then simulated according to the following two different input power densities: one simulation with a power density of $1 \times 10^{10}$ W/m$^3$ with a high power pulsed source, and a series of simulations at a power density of $1 \times 10^9$ W/m$^3$ to represent the power density in a 30 kW source microwave with 2.45 GHz applicator. The authors found that, at $1 \times 10^9$ W/m$^3$ input energy and 0.01s microwave exposure, the amount of grain boundary damage was less than 50%. However, at a higher value of $1 \times 10^{10}$ W/m$^3$ and lower exposure of 0.001s, the grain boundary damage increased to 74.3%, which confirms that the influence of microwave power density on the degree of ore liberation is high. The practical implication of the results presented in Ali and Bradshaw's [120] study is that for ores that are less amenable to continuous wave microwaves with low power densities, very high power pulsed microwave systems could be used for an economical treatment. The authors further extended their investigations to characterize the amount of microcracks induced in a conceptual binary ore consisting of 10% galena and 90% calcite by microwaves and to study the effect of applied power density and ore texture on the quantity of microcracks by using a bonded-particle modeling approach [55]. The results of this study demonstrated that through the implementation of higher microwave power density, more microcracks could be induced at the same energy input. A pulsation mode of microwaves also helps to liberate the absorbent minerals of the original size. Ultimately, the study shows that for the treatment of fine-grained ores at economical energy inputs, a higher power density is needed. Later, in another publication, Bradshaw et al. [121] showed that with finer textured mineral ores, the required specific energy input had to be greater to produce the same amount of damage. For the same power density and exposure time of the applied microwave, smaller thermal stresses generate a finer textured ore.

In 2010, using COMSOL multiphysics software a numerical modeling of microwave heating of an insulated sphere with a radius of 3 cm was presented by Lovás et al. [74]. The aim of this study was to investigate the effects of microwave irradiation on temperature distribution in the samples of andesite, siderite, magnesite, chalcopyrite, and pyrite. From the results of the models, the authors concluded that the response of a mineral when it is subjected to a microwave treatment depends greatly on the material's electromagnetic and thermal properties, such as its dielectric properties, specific heat capacity, and thermal conductivity. Therefore, Lovás et al.'s work verified that different material properties of a mineral have a great impact in the heating and breakage processes of the rock under microwave irradiation. To make this section more concise, a summary list with technical details of other published numerical works that were found in the literature from 2010 to 2020 is provided in Table 4.

**Table 4.** Publications on numerical simulations of microwave-assisted heating and pre-conditioning of rocks and minerals that were found through a chronological literature review (2010-2020)

| Numerical methods, modeling approach, software | Microwave type and geometry, MW input numerical parameters | Rock/mineral name, Sample input model parameters | Main findings and highlights |
|---|---|---|---|
| | | Peng et al. [122] | |
| FDM, coupled 2D electromagnetic and thermal model, MATHEMATICA 7.0 | N/A, MW power densities (0.5-4 MW/m$^2$) for 60s exposure at 0.915 GHz; and power density (1 MW/m$^2$) for various exposure times (1, 60, 300, and 600s) at 2.45 GHz | Coal, homogeneous magnetite block with a varying dimension of 2L x 2L m (L=0.2, 0.15, 0.1, and 0.05 m) | Microwave heating of coal at 0.915 GHz exhibited better heating uniformity than the same irradiation at 2.45 GHz because of the larger penetration depth of the MW. Under the same MW conditions applied to the same material, a MW irradiation with 2.45 GHz would dissipate in the area closer to the surface than that at 0.915 GHz. |
| | | Hartlieb et al. [89] | |
| FEM, 3D thermal and thermomechanical model, ABAQUS v6.10 | Multi-mode; MW power level (3.2 kW), applied power density of 16.8×10$^6$ W/m$^3$ at 2.45 GHz; and exposure times (up to 900s) | Basalt, cylindrical (h=50 mm, r=25 mm) sample; various temperature dependent values of $k$ (W/mK), $c_p$ (J/kgK), $E$ (GPa), and $v$ (1) applied in FE simulations | For a slow MW absorption process (when the heating rate is small compared to the heat transfer rate in grains and between the sample and its surrounding), a fine-grained rock like basalt does not show effects of differential heating and thermal expansion of individual grains. |
| | | Jokovic [123] | |
| FDTD, 3D electromagnetic and thermal model, QuickWave 3D | Multi-mode; Pentagonal and rectangular cavity, MW power levels (15, 20 kW), at various frequencies (2.45 and 0.915 GHz); and various exposure times | various ore particles, e.g. including monzonite ore, modeled with different dielectric properties applied to the rock models | Rectangular cavity provided more uniform and efficient heating than the pentagonal cavity. The study shows although multi-mode MW heating should provide, in the best case, uniform heating throughout the cavity, optimization of feed positions must be completed for maximum efficiency. |
| | | Wang and Djordjevic [124] | |
| FEM, 2D electromagnetic, thermal and mechanical model, ANSYS | N/A, MW power densities (10$^{10}$, 10$^{11}$ W/m$^3$); frequency (2.4 GHz); and exposure times (0.00043s, 0.00044s, 0.00045s, 0.00046s) | a single disc-shaped grain of pyrite surrounded by a larger disc of calcite, various particles grain sizes for the ores | The size and thermal properties of the rock under microwave exposure can significantly affect thermal heating and subsequent thermal stress. The study suggests that a high-power density combined with a short heating interval offers the best energy efficiency. |
| | | Meisels et al. [125] | |
| FDTD, 2D electromagnetic, thermal and mechanical model, ABAQUS v6.12 | Single-mode, MW power level 25 kW and assumed losses of 30% resulting in a power input of 17.5 kW | Basalt, Gabbro and Granite, a block (15 × 15 × 20 cm$^3$) rock sample at 30 cm distance from MW from antenna | Different morphologies resulted in a change to the maximum stress and the initiation of cracks. Temperature gradients and induced stresses will occur on short distances. |
| | | Charikinya [126] | |
| DEM, (a) 2D and (b) 3D micromechanical and thermal models, PFC2D/PFC3D | Single-mode, MW power density (10$^{10}$ W/m$^3$) representing a power density in a 6 kW power cavity with a frequency of 2.45 GHz; and various exposure times (0.01-4s) | Massive sulfide ores, (a) 2D rectangular specimen (W15 × L20 mm$^2$) (b) 3D parallelepiped specimen (15 × 15 × 20 mm$^3$) | The results from 2D models suggest that model resolution had a great impact on the magnitude of simulated crack damage. From 3D models, the author found that the majority of the cracks observed were tensile cracks with very few shear cracks. |
| | | Hong et al. [127] | |
| FEM, coupled 3D electromagnetic and thermal model, COMSOL Multiphysics | Multi-mode; cavity (W267×L270×H188 mm$^3$), MW power levels (0.5-3 kW); frequencies (2.4-2.5 GHz); and exposure times (up to 300s) | Coal, a cylindrical (h = 50 mm, r = 12.5 mm) sample | MW heating behavior of the coal samples was highly affected by the samples' positioning in the cavity and the MW frequency and power level. |
| | | Lin et al. [128] | |
| FEM, coupled 3D electromagnetic and thermal model, COMSOL Multiphysics | Multi-mode; cavity (W630×L650×H660 mm$^3$), MW power levels (0.5-6 kW); frequencies (2.4-2.5 GHz); and exposure times (0-600s) | Coal, cylindrical (h = 60-100 mm, d = 50 mm) samples; various samples' permittivities (dielectric constants and loss factors) applied | (1) Better thermal heterogeneity of coal was obtained with larger MW power levels, (2) coal sample temperature increased while thermal heterogeneity decreased with loss factor, and (3) efficient heating was found to be achieved at an optimal frequency of 2.45 GHz. |

*Note: In the table the term "microwave" is abbreviated as MW.

**Table 4.** continued

| Numerical methods, modeling approach, software | Microwave type and geometry, MW input numerical parameters | Rock/mineral name, Sample input model parameters | Main findings and highlights |
|---|---|---|---|
| | | Huang et al. [129] | |
| FEM, coupled 3D: (a) electromagnetic, heat and mass transfer model; (b) electromagnetic, thermal and mechanical model, COMSOL Multiphysics | Multi-mode; cavity (W267×L270×H188 mm$^3$): (a) MW power levels (0.5-2 kW); frequencies (1.95-3.7 GHz); specific moisture capacity (0.5-10%); and exposure times (0-60s); (b) MW power level 500 W; frequency 2.45 GHz; and exposure times (0-300s) | Coal, a cylindrical sample (h = 60 mm, r = 25 mm); various samples' permittivities (dielectric constants and loss factors) applied | (a) Microwave heating of a coal sample was responsive to input MW power level and frequency; and the best MW heating effect was found in the coal with low water saturation. (b) The results showed that the average permeability of coal rose by 2.2 times after 300s MW exposure with 500 W input power level at the frequency of 2.45 GHz. |
| | | Li et al. [130] | |
| FEM, coupled 3D electromagnetic and thermal model, COMSOL Multiphysics | Multi-mode; cavity (W630×L650×H660 mm$^3$), MW power level 1 kW at 2.45 GHz frequency; and exposure times (10-300s) | Coal, cylindrical sample (h = 100 mm, d = 50 mm) | Variation in electric field norms was observed because of nonuniform electromagnetic distribution that resulted in a selective heating of the coal sample. When water evaporation was included, temperature increased nonlinearly. |
| | | Li et al. [131] | |
| FEM, coupled 3D electromagnetic, thermal and mechanical model, COMSOL Multiphysics & AutoCAD | Multi-mode, MW power level 3 kW applied at 2.45 GHz frequency; for 120s exposure time | Pegmatite, cylindrical samples (h = 100 mm, d = 50 mm); a thin section (0.34 × 0.47 in$^2$) sliced from the center of the rock for 2D analysis | Dielectric constants, coefficients of thermal expansion, and sizes of mineralogical boundaries in minerals have a significant impact on the effects of MW treatment. By increasing MW exposure time, the rate of generation of von Mises stresses increases rapidly along the interfaces between different minerals. All the stress-strain curves showed elastic deformation behavior. |
| | | Li et al. [132] | |
| FEM, coupled 3D electromagnetic and thermal model, COMSOL Multiphysics | Multi-mode; cavity (W630×L650×H660 mm$^3$), MW power levels (1-6 kW); at 2.45 GHz frequency; and exposure times (60s, 120s, 180s, 240s, 300s) | Coal, cylindrical sample (h = 30 mm, d = 25 mm) | By increasing the input MW power level, the electric field intensity and temperature of coal and the spatial heterogeneity of the electric and thermal fields all increased. MW heating can induce interconnection of pores and fractures in coal. |
| | | Jinxin et al. [133] | |
| FEM, coupled 3D electromagnetic, thermal, and mechanical model, COMSOL Multiphysics | Multi-mode; cavity (W267×L270×H188 mm$^3$), MW power level 500 W; at 2.45 GHz frequency; and exposure times (0-300s) | Coal, a cylindrical sample (h = 60 mm, r = 25 mm) | The average permeability of coal increased from 1.65×10$^{-16}$ m$^2$ to 3.63×10$^{-16}$ m$^2$ after MW irradiation of 500 W at 2.45 GHz and 300s exposure. |
| | | Yuan and Xu [134] | |
| FEM, coupled 3D electromagnetic, thermal, and mechanical model, MATLAB & COMSOL Multiphysics | Multi-mode; cavity (W267×L270×H188 mm$^3$), MW power level 3 kW; at 2.45 GHz frequency; and exposure times (10-90s) | Basalt, a cylindrical sample (h = 100 mm, d = 50 mm) | The uneven distribution of temperature caused damage because of the different thermal stresses produced in the rock sample. Tensile failure at the surface of rock was observed. Finally, radial cracks were observed from both the numerical models and the experiments. |
| | | Zhang et al. [135] | |
| FEM, coupled 2D electromagnetic, thermal, and mechanical model, COMSOL Multiphysics | Single-mode; cavity (W300×L500 mm$^2$), MW power level 100 kW; at 2.45 GHz frequency; and 45s exposure time | Granite & Limestone, rectangular samples (W300mm) at various distances (12, 14, 16 cm) from MW from antenna | The distance from MW antenna has a great influence on the electromagnetic field, temperature, mechanical stress, and plastic zone distribution in rock. At the same distance, limestone showed higher electromagnetic field intensity than granite; instead, the rock surface temperature of granite was higher. |
| | | Hidayat et al. [136] | |
| FEM, coupled 3D electromagnetic and thermal model, ANSYS v17 | Multi-mode, MW power levels (0-3 kW & 5.55 kW); at 2.45 GHz frequency; and exposure times (up to 240s) | Ilmenite, cylindrical samples (r = 0.035 m, L = 0.015, 0.03, 0.045, and 0.06 m) slab samples (L0.07 × W0.07 m) and various thickness (b = 0.015, 0.03, 0.045, 0.06 m) | Hotspot location in ilmenite changed as the thickness of sample varied. Higher temperature distribution was observed in thinner samples. |

*Note: In the table the term "microwave" is abbreviated as MW.

## 4. Technical and economic analyses

This section aims to give an overview of the published studies related to energy-related (consumption, savings, etc.) investigations and to present techno-economic analyses on the results of microwave irradiation of rocks and minerals discussed in earlier sections, for future possible implementation of a pilot-scale microwave in industrial usage. As discussed by Napier-Munn et al. [137], the mechanical size reduction of solids (comminution) is one of the most energy intensive processes for mineral processing plants. Thus, there have been serious efforts in the past two decades for improving the efficiency of comminution processes. Through investigations in the findings of experimental research studies, it can be confirmed that reduced comminution energy and improved recovery processes are practically achievable at economically viable microwave energy inputs [75,138]. However, depending upon which types of microwave applicators are used, single-mode or multi-mode, the outcome varies significantly. For example, a single-mode microwave cavity provides relatively high power densities in the absorbing phases of minerals. It was found that 1 kg batches of copper carbonatite ore treated at 15 kW and 0.1s microwave exposure (or 0.4 kWh/t) showed a reduction in strength of over 50%; and little improvement was achieved for longer microwave exposures as discussed by Kingman et al. [75]. The copper ore recovery after microwave treatment was found to be increased by 85% to 89.5% [139]. In addition, comparative batch flotation experiments by Sahyoun et al. [140] on copper ore samples microwaved with power levels of 5-12 kW at 0.1-0.5s exposure times demonstrated that improvements in copper recovery of between 6-15% could be achieved when the results of treated ores were compared with untreated samples. On the other hand, several studies on microwave irradiation of minerals in multi-mode cavities have shown promising results. For instance, after microwave treatments of 20 grams chalcopyrite in a multi-mode microwave cavity with 3 kW power level and the frequency of 2.45 GHz for 0, 5, 10, and 20s of microwave exposures, Da Silva et al. [141] found that microwave treatment increases the mineral's specific surface area and porosity and changes the pore size distribution. Therefore, the authors concluded that chalcopyrite's leaching kinetics process could be improved by multi-mode microwave irradiation. The same conclusion was also drawn in a study by Zhu et al. [142] in which the effects of microwave heating parameters on the pore structure of oil shale samples were evaluated. In conclusion, these results verify that today microwave treatment of ores can be done economically. However, depending upon the objective of the project, a wise selection between the two different microwave applicators, single-mode or multi-mode, should be considered.

**Table 5.** Results of microwave treatment effects on strength reductions in rocks and minerals: a comparison between the amount of microwave energy input and the mechanical changes in different rock samples

| Reference | Type of rock/mineral | MW operating parameters and type (M or S*) | Result (UCS, Cracking, energy) |
|---|---|---|---|
| Vorster et al. [25] | Massive copper ore | P = 2.6 kW, t = 90s, **M** | BWI of the ore reduced up to 70% |
| Kingman et al. [75] | Copper carbonite ore | P = 15 kW (0.83 kWh/t). t = 0.2s, **S** | 30% reduction in impact breakage parameters |
| Kingman et al. [138] | Lead-zinc ore | P = 10 kW, t = 0.1s, **S** | Up to 50% reductions in strength of the ore |
| Charikinya [126] | Sulfide ore | Input MW energies (2.11-2.65 kWh/t), **S** | Cracks volume increased by 500% |
| Kumar [84] | Iron ore | P = 900 W, t = up to 120s, **M** | Specific rate of breakage increased by an average of 50% |
| Wang & Forssberg [77] | Quartz and limestone | P = 7 kW, t = 10 min, **M** | UCS decreased from 50 MPa to 25 MPa and 40 MPa to 35 MPa for quartz and limestone |
| Satish et al. [79] | Basalt | Power density 1 W/g, t = 360s, **M** | 42% increase in penetration of cutter |
| Hassani et al. [4] | Basalt | P = 5 kW, t = 65s, **M** | 30% reduction in uniaxial compressive strength |
| Sikong [82] | Granite | (a) P = 850 W, t = 30 min; (b) P = 600 W, t = 10 min (quenched sample), **M** | (a) 60% reduction in samples' compressive strength (b) 38% reduction in cutting rate |
| Zeng [108] | Granite | P = 1.4 kW, **M** | UCS of granite decreased from 88.17 MPa at 25 °C to 18.61 MPa at 800 °C |
| Kobusheshe [83] | Kimberlite | Input MW energy of (a) 9 kWh/t; and (b) 6.81 kWh/t, **S & M** | (a) 40% reduction in point load strength and (b) 10% reduction in the mean UPV |
| Singh et al. [100] | (a) Coal sample; and (b) Manganese ore | (a) P = 180 W; (b) P = 900 W, t = 1, 3, 5 min, **M** | (a, b) 17.1% average increase in carbon recovery for coal |
| Zheng [102] | Gabbro | P = 2 kW, t = 30-120s, **S** | 55% reduction in the overall p-wave velocity |

*Note: In the table the term "microwave" is abbreviated as MW; and the two MW applicator types of single-mode and multi-mode are denoted by **S** and **M**, respectively.

For further technical investigation on the reported experimental data discussed in the literature review, Table 5 is given as an overview of the amounts of rock/ore strength reduction from different measurement techniques with respect to different microwave input operating parameters. It can be seen that the amount of microwave energy input has a direct relation to the strength reduction of rocks and minerals. The feasibility and economical potential of microwave irradiation of rocks and minerals was demonstrated after a techno-economic study by Bradshaw and colleagues [36] looked into the potential profitability of microwave-assisted comminution and liberation of mineral ores. The preliminary economic analyses presented in this study showed that the overall cost of the microwave equipment was found to be in a range between $0.16 US to $0.85 US per ton of ore. In a detailed investigation, the study tabulated the estimated data of operating and cost parameters for industrial scale applicators for two appropriate ISM frequencies, 433 and 915 MHz (see Table 6). It should be noted that the estimated values were only provided for a Constant Wave (CW) operation and do not involve pulsed or other types of microwaves.

**Table 6.** Estimated operating and cost parameters for industrial scale applicators (table modified from Bradshaw et al. [36])

| Frequency, MHz | 433 | | 915 | |
|---|---|---|---|---|
| | Most expensive scenario | Least expensive scenario | Most expensive scenario | Least expensive scenario |
| Power kW | 500 | | 100 | |
| Estimated feed top size (mm) | 90 | | 50 | |
| Cross section of applicator m$^2$ | 0.32 | | 0.07 | |
| Critical energy density Jm$^{-3}_{abs}$ | 1×10$^8$ | 0.5×10$^8$ | 1×10$^8$ | |
| Microwave energy consumption kWh/t | 1.11 | 0.55 | 1.11 | |
| Order of magnitude unit capital cost US$/kW | 20,000 | 10,000 | 7,000 | 4,000 |
| Electricity cost US$/kWh | 0.035 | | 0.035 | |
| Annual operating hours | 8,150 | | 8,150 | |
| Order of magnitude capital cost US$ | 10,000,000 | 5,000,000 | 700,000 | 400,000 |
| Amortized cost US$/yr | 2,983,155 | 996,260 | 208,820 | 79,700 |
| **Overall cost US$/t** | **0.85** | **0.16** | **0.14** | **0.06** |

In Table 6, the estimated capacity, cost, and other operating parameters for the most appropriate ISM frequencies, 433 and 915 MHz, for industrial scale operation are given. It is clearly estimated that for large-scale operations, an appropriate capacity can be achieved at 433 MHz frequency, while for smaller operations, parallel units at 915 MHz might be feasible. Overall, Bradshaw et al.'s [36] study estimated that the overall cost of microwave equipment was found to vary in a range from lower bound ($0.06 US/t) to upper bound ($0.85 US/t) on the total amortized capital and operating expenditure. This study initiated for the first time a concluding remark for possible future industrial implementation of an economic microwave usage for rock breakage and mineral processing industries.

As discussed earlier, the high consumption of energy in the process of microwave treatment of rocks and ores before their processing is the primary challenge that must be appropriately addressed before the technology becomes feasible and economical. Assuming that microwave treatment reduces 50% of the UCS of the rocks and $0.28/kWh to be the cost of electricity, $8.66/t is the electricity cost of implementation of the microwave treatment system working with 0.76 UCS reduction per energy rate (i.e. MPa per kWhr/m$^3$). However, by increasing this rate to ten times its initial value, the electricity cost of the microwave treatment system can be reduced to $0.89/t, which is much smaller in comparison to the total mining cost, e.g. higher

economic feasibility. Therefore, in order to achieve these higher rates, higher power densities are required. But there are other researchers, like Didenko et al. [76], who used $1.71 \times 10^5$ kW/m$^3$ power densities; however, these power densities led to low UCS reduction per energy rates. Thus, to investigate feasibility of microwaves in the reduction of the overall rocks' strength, many parameters, including the role of material properties' variation, input microwave operating parameters, experimental conditions, etc. should be accurately determined. As indicated in almost all recent research studies on microwave irradiation of rocks and minerals, the key to techno-economically feasible implementation of any microwave-assisted rock treatment system for the industry is to reduce energy intensity effectively; and therefore, by selecting an optimal microwave power density, a feasible design of microwave-rock pre-conditioning systems can be achieved.

## 5. Conclusion

In conclusion, the present review study demonstrates that the application of microwaves in the rock breakage industry, in both mining and processing of rocks and minerals, started after subsequent studies confirmed that microwaves are more competitive than any other assistive rock breakage techniques, considering:

- Microwaves heat up rocks on the basis of their dielectric properties; therefore, energy is not wasted to warm up the whole rock body, since rocks are composed of various minerals with different dielectric properties [143],
- Microwaves are faster as they rely on selective heating of minerals (some minerals have a rapid response to the applied electromagnetic waves) [144],
- Microwaves have high a level of safety and automation [145].

Taking all the discussed literature into account, several experimental studies have been carried out in recent years to investigate and address the effects of microwave irradiation on various types of rocks and minerals. These studies were able to successfully identify the importance of microwave energy as a prospective application in future rock breakage operations and mineral processing applications in the civil and mining industries. In addition, in tandem with experimental research, computational studies have come to play an important role in understanding how different mechanisms are involved in microwave-assisted rock pre-conditioning and breakage processes. The development of a numerical model is considered an enhancement of the understanding of microwave-assisted rock pre-conditioning and breakage and thermomechanical characteristics of rocks after their exposure to microwaves at different microwave operating parameters, either in single-mode or multi-mode microwave cavity systems. As discussed and reviewed in this paper, today, the vital role of computer simulations in the field of mining and civil engineering is recognized for researchers and even the industry. Serious efforts have been made by researchers for developing a reliable and trustworthy numerical model that mimic the correct electromagnetic wave propagation in the multi-mode microwave system. However, although several studies have been done in recent years to investigate the effects of multi-mode microwave irradiation on rocks with the numerical modeling approach (e.g. finite element method), no numerical model of multi-mode microwave irradiation was found that: (1) is fully coupled for electrical and thermal multiphysics interactions, (2) represents a correct electromagnetic wave propagation inside the multi-mode microwave cavity as in the experiment, and more importantly (3) is validated against experimental data of microwave irradiation tests on various rock types with different microwave settings.

**Acknowledgement**

Financial support from the McGill Engineering Doctoral Awards (MEDA) and the Dr. Gerald G. Hatch and Lorne Trottier Graduate Fellowships in Engineering is gratefully acknowledged.